\documentclass[aps,prb,twocolumn,amsmath,amssymb,superscriptaddress,10pt]{revtex4-2}
\usepackage{fix-cm}
\usepackage	[pdftex, bookmarks, colorlinks, breaklinks]							{hyperref}

\usepackage				{xr}						
\usepackage	[english]		{babel}
\usepackage	[export]		{adjustbox}
\usepackage	[caption=false]	{subfig}
\usepackage				{graphicx, epstopdf, epsfig, rotating, floatpag}
\usepackage				{xcolor, color, verbatim, varioref, units, cancel}
\usepackage				{booktabs, tabu, longtable, dcolumn}
\usepackage				{amsmath, amssymb, amsfonts, nicefrac, xfrac, gensymb, braket, bm}
\usepackage				{paralist, enumerate, lipsum}
\usepackage				{dsfont}					
\usepackage{enumerate}

\newcommand{\SPC}{structural percolation cluster}

\usepackage{newtxtext}
\usepackage{newtxmath}

\usepackage{lipsum}

\hypersetup				{colorlinks=true, linkcolor=black, citecolor=blue, filecolor=dullmagenta, urlcolor=blue}	
\usepackage[sectionbib]{bibunits}
\defaultbibliography{ref}
\defaultbibliographystyle{apsrev4-2}

\begin{document}

\title{
    Hydrodynamic Memory and Long-Time Tails in Clean Frustrated Magnets
}

\author{Yufei Pei}
\affiliation{TCM Group, Cavendish Laboratory, JJ Thomson Avenue, Cambridge CB3 0HE, United Kingdom}
\address{Max Planck Institute for the Physics of Complex Systems, 01187 Dresden, Germany}

\author{Claudio Castelnovo}
\affiliation{TCM Group, Cavendish Laboratory, JJ Thomson Avenue, Cambridge CB3 0HE, United Kingdom}

\author{Roderich Moessner}
\address{Max Planck Institute for the Physics of Complex Systems, 01187 Dresden, Germany}

\begin{abstract} 
In a simple, clean but constrained magnet, we identify a self-interacting random walk with memory. For the motion of a single monopole -- a fractionalized quasiparticle in spin ice -- this produces a subtle and unusually slow relaxation toward diffusive motion. This is manifested as an algebraic long-time tail in the velocity autocorrelation function, decaying as $t^{-3/2}$. At finite monopole density, the interactions between the trails of different monopoles introduce an additional timescale, corresponding to the disruption of a monopole's memory by other monopoles, and leading to an exponential cutoff of the long-time tail. Our results identify clean frustrated magnets as microscopic platforms for studying self-interacting stochastic processes, hydrodynamic memory, and the emergence of non-Markovian quasiparticle transport from local Markovian dynamics.
\end{abstract}

\maketitle
%
%

\emph{Introduction ---} 
The transition from microscopic stochastic steps to macroscopic transport is a central pillar of statistical mechanics~\cite{haus1987diffusion, kubo2012statistical}. For ordinary random walks, the memory of individual steps is rapidly lost, giving rise at large scale to diffusion governed by only a few effective transport coefficients. Many physical stochastic processes, however, are not of this memoryless kind: their future evolution depends on the history written by earlier motion, either through the territory already explored or through local degrees of freedom modified along the trajectory~\cite{foster2009reinforced, barbier2022self}. Such stochastic processes with memory include self-interacting and reinforced random walks, where the walker responds to its own footprint, and they provide a natural language for describing transport in media that are themselves dynamically rearranged by the moving object~\cite{robin2007survey}.

Frustrated magnets provide a natural many-body setting in which such memory effects can arise. Their low-energy states are often governed by local constraints rather than conventional long-range order, giving rise to spin-liquid regimes, emergent gauge structures, and fractionalized or topological quasiparticles~\cite{balents2010spin, savary2016quantum, Knolle_2019}. More broadly, the dynamics and transport of magnetic quasiparticles are central themes across the field, ranging from magnons in ordered magnets~\cite{cornelissen2016magnon}, to spinons and Majorana excitations in quantum spin liquids~\cite{savary2016quantum, knolle2014dynamics, Knolle_2019}, skyrmions in chiral magnets~\cite{Nagaosa2013topological}, and magnetic monopoles in spin ice~\cite{castelnovo2008magnetic, castelnovo2012spin}. In constrained frustrated magnets, however, the motion of an excitation can do more than carry spin, energy, or topological charge: it rearranges the spin background, while the rearranged background constrains its subsequent motion~\cite{castelnovo2008magnetic, jaubert2009signature, castelnovo2012spin,hallen2022dynamical}. The spin configuration therefore acts both as a dynamical medium and as a memory register. This makes constrained frustrated magnets a natural bridge between stochastic processes with memory and quasiparticle transport in correlated magnetic matter.

In this work, we study this feedback mechanism in the nearest-neighbor classical spin-ice model, using it as a minimal model of a disorder-free constrained magnet with mobile emergent defects. Its low-energy configurations obey a local ice rule, and violations of this constraint behave as deconfined magnetic monopoles moving on the dual diamond lattice. Under local single-spin-flip stochastic dynamics, each monopole hop flips a spin, thereby both moving the defect and rewriting the background that constrains its future motion. The resulting Dirac-string trail is neither quenched disorder nor an externally imposed memory field, but part of the evolving spin configuration itself. Thus, although the full spin dynamics is Markovian, the projected monopole motion realizes an effective random walk with self-generated memory in a clean frustrated magnet.

We show that this clean constrained magnet realizes a self-interacting stochastic process with a hydrodynamic memory tail. Tracking the motion of an isolated emergent defect, we find that its velocity autocorrelation ($C(t)$ in the following) decays algebraically over long times, yielding a surprisingly slow approach to ordinary diffusion. Such algebraically decaying correlations are known as long-time tails~\cite{alder1970decay,paul1981observation,fox1983long,naitoh1990long,ferrario1997long}. In Brownian diffusion, their $C(t)\sim t^{-d/2}$ scaling arises from memory effects of the surrounding fluid on a moving particle~\cite{alder1970decay}. In our (three-dimensional) case, the same scaling~\cite{Note2} originates instead from memory written locally onto the constrained spin background. This behavior is not reproduced by a random walk in a static constrained environment, but is captured by a self-attractive random walk in which the walker interacts with permanent memory along its own trajectory. At finite defect density, the motion of other defects progressively erases this memory, introducing a density-controlled exponential cutoff to the algebraic tail. Our results therefore identify frustrated magnets as microscopic, disorder-free platforms for studying how non-Markovian transport and memory erasure emerge from local Markovian dynamics. 
%
%

\begin{figure}[t!]
    \centering
    \begin{minipage}[c]{0.48\linewidth}
        \includegraphics[width=\linewidth]{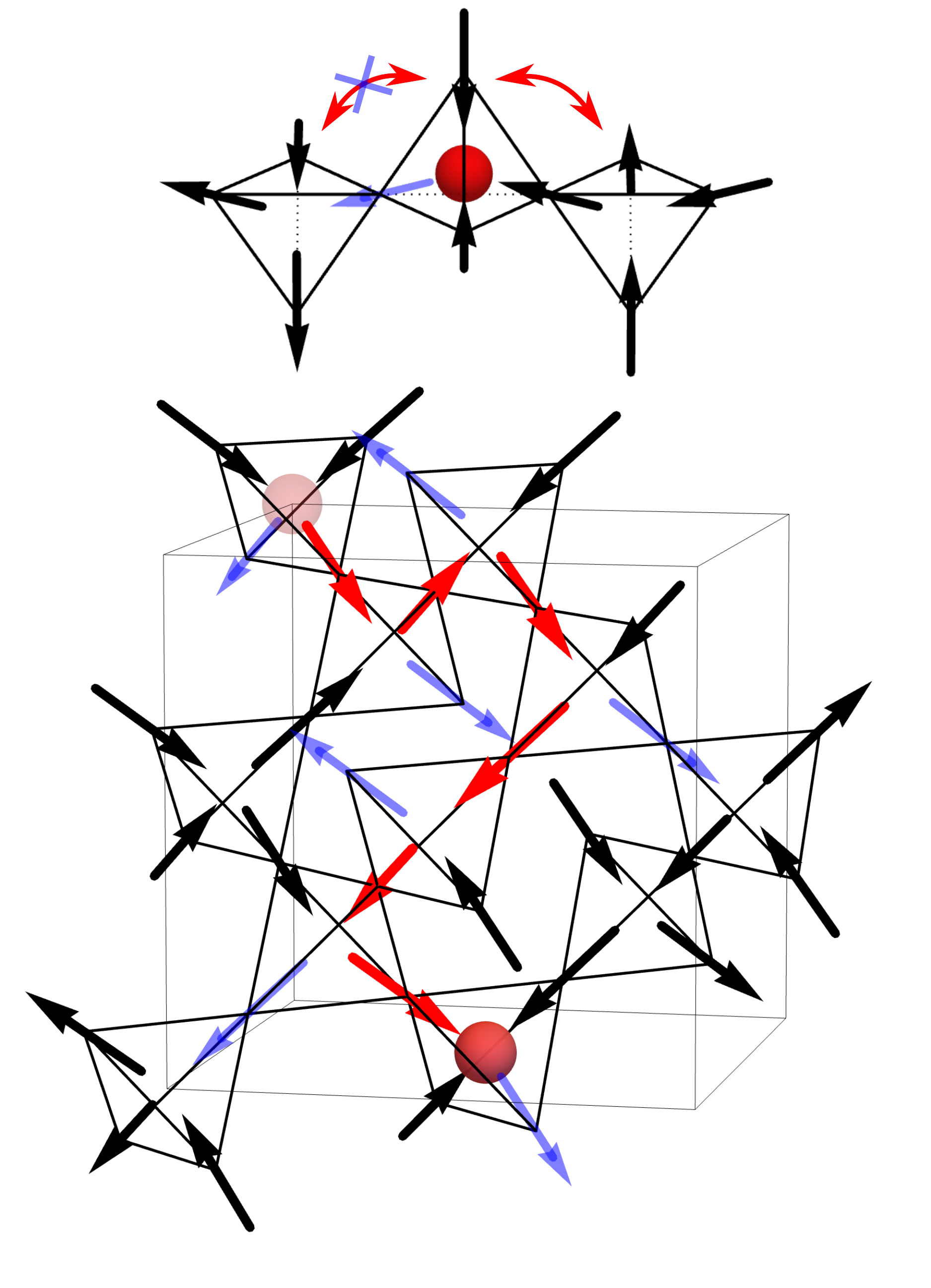}
        \vspace{1em}
    \end{minipage}
    \begin{minipage}[c]{0.48\linewidth}
        \includegraphics[width=\linewidth]{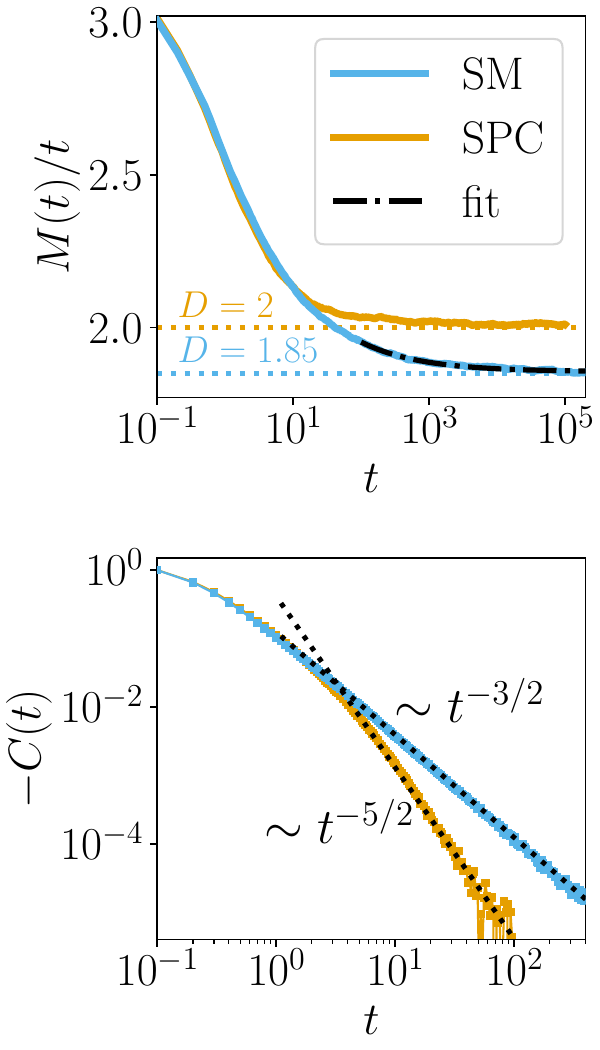}
    \end{minipage}
    \caption{\label{fig:1} Schematic illustration of spin ice and monopole hopping. Top left panel: The middle tetrahedron ($3$-in-$1$-out) can hop to the right without increasing the system energy, by flipping a majority spin; hopping to the left instead creates a costly double monopole, by flipping a minority spin, and is suppressed at low temperatures. Bottom left panel: example of a (zero-energy) monopole trail, with the flipped spins shown in red, and the blocked spins at each step shown in blue. The cubic box indicates one unit cell of $16$ spins. Right panels: mean-square displacement $M(t)$ (top) and (negative of the) velocity autocorrelation function $C(t)$ (bottom), for the standard model monopole dynamics (SM; system size $L=180$, see App.~\ref{app:methods}) and for the {\SPC} (SPC, see main text) random walk. The MSD of the two processes are similar at short times, but the {\SPC} walk relaxes more quickly to the diffusive limit than the monopole random walk. We also plot the fit $M(t)=Dt+A_1t^{1/2}$ to the long-time behavior of the monopole random walk ($D\simeq1.85\pm0.01$ and $A_1\simeq0.9\pm0.1$). 
    The VACFs for the two random walks exhibit distinct asymptotic behaviors at large times, with the {\SPC} scaling as $t^{-5/2}$ (consistently with previous studies~\cite{van1982transport}) and the monopole walk scaling as $t^{-3/2}$ (hydrodynamic behavior).}
\end{figure}

\emph{Model and dynamics ---} 
We consider nearest-neighbor classical spin ice (nnSI)~\cite{udagawa2021spin}: 
\begin{equation}\label{2:H}
    \mathcal{H}=J\sum_{\langle i, j\rangle}\,S_iS_j
    \, , 
\end{equation}
where $S_i = \pm 1$ are classical Ising spins on the sites $i$ of a pyrochlore lattice of size $16L^3$. We set $J=1$ as our reference energy scale. The ground states obey the $2$-in-$2$-out ice rule, whereby each tetrahedron has two spins pointing oppositely to the other two, as illustrated in Fig.~\ref{fig:1}. Local violations of the ice rule create `magnetic monopole' excitations. At lowest energy, these are $1$-in-$3$-out or $3$-in-$1$-out tetrahedra, which behave as deconfined point-like topological quasiparticles~\cite{castelnovo2012spin}, charged under the emergent U(1) gauge symmetry and costing each an energy $\Delta = 2$. We focus on the low-temperature regime, where the monopole density is exponentially suppressed ($\rho \sim e^{-\Delta/T}$, setting $k_B=1$), and double monopoles ($4$-in and $4$-out tetrahedra) have vanishing density and can be safely neglected. In this regime, nnSI can be modeled as a dilute gas of monopoles in an otherwise pristine ice-rule background. 

Attempts to model monopole dynamics in spin ice began with the seminal work of Ryzhkin~\cite{ryzhkin2005magnetic}, followed by Jaubert and Holdsworth's single-spin flip stochastic dynamics, later termed the `standard model'~\cite{jaubert2009signature}. Under such dynamics, each spin attempts to flip stochastically with a characteristic timescale (set to unity throughout this work), and with Metropolis-Hastings acceptance probability. 
Other models of spin ice dynamics have been proposed, most notably the recent `beyond the standard model' dynamics~\cite{hallen2022dynamical}, which we discuss for completeness in App.~\ref{app:bSM} and~\ref{app:psd}. 

Monopoles are created and annihilated in pairs, and are able to wander across the system at no further energy cost. Successive spin flips thereby generate random walk trails of monopoles on the dual diamond lattice, leading to a reaction-diffusion description of spin ice dynamics. 
Monopole motion changes the background, and the background dictates how monopoles move at the microscopic scale -- leading to a close interplay that is the root cause of the interesting phenomenology uncovered below. 

In our work, we focus primarily on the random walk of isolated monopoles (relevant at low temperatures). 
As shown in Fig.~\ref{fig:1} (top left panel), a key feature of such random walks is that, at each step, only $3$ out of the $4$ possible directions (i.e., the majority spins) are allowed to flip. The minority spin direction is effectively blocked by its large energy cost (flipping it would produce a double monopole). The path traversed by a monopole is necessarily a chain of head-to-tail aligned spins (termed a `Dirac string'~\cite{castelnovo2008magnetic, castelnovo2012spin}), and a monopole can always backtrack along it (see the bottom left panel in Fig.~\ref{fig:1}). 
The blocked directions are an important difference between monopole motion and ordinary random walk on the diamond lattice. Another, more subtle difference derives from the power-law correlations between the spins due to the emergent U(1) gauge symmetry. The walk is therefore self-interacting and history-dependent, with motion at time $t$ correlated with its entire prior trajectory. 
%
%

\emph{Single-monopole random walk ---} 
We compute the mean-square displacement (MSD) of a monopole random walk, 
\begin{equation}\label{3:MSD}
    M(t)\equiv \left\langle\left[\mathbf{r}(t)-\mathbf{r}(0)\right]^2\right\rangle
    \, , 
\end{equation}
where we set our unit length to be the distance between the centers of adjacent tetrahedra. In the case of ordinary random walk on the diamond lattice (no blocked directions nor spin correlations), one expects asymptotically $M(t) = 4t$. If the blocked directions are entirely uncorrelated (and re-drawn at each time step randomly), one expects $M(t) = 3t$. 

The behavior for our monopole random walk is shown in the top-right panel of Fig.~\ref{fig:1}. We find that $M(t)/t$ tends to $3$ at short times, as expected from the $3$ available directions of motion at each step. 
At long times however, the diffusion constant decreases substantially, as a result of the fact that backtracking is always possible. We obtain $D=\lim_{t\to\infty}M(t)/t\simeq1.85\pm0.01$. 
Note the long times required to approach the asymptotic value ($\gtrsim 10^5$ for a system of size $L=180$, see App.~\ref{app:methods}). 

We compare the monopole random walk result to the case of a random walk on the so-called structured percolation cluster~\cite{nilsson2023dynamics} (SPC) obtained by considering a dimer covering of the diamond lattice, and then removing all the edges covered by a dimer ($1/4$ of all the edges). Similarly to spin ice, a particle on a {\SPC} can only hop across $3$ of the $4$ possible directions at each site. 
The resulting MSD reproduces that of the monopole random walk at short times, while featuring a faster convergence to the diffusive limit, as shown in Fig.~\ref{fig:1} (top right panel). 
A significant difference between the monopole random walk and the {\SPC} walk begins to be noticeable when the random walk trajectories form loops. While there are no constraints about traversing a loop in the {\SPC} walk, a monopole cannot traverse the same spin twice in the same direction, which acts to suppress loop dynamics. The smallest loops on the diamond lattice have length ${l_{min}} =6$, and we correspondingly expect a departure between the behavior of the two models around $t \sim l_{min}^2$, consistent with the top right panel in Fig.~\ref{fig:1}. 
%
%

\emph{Hydrodynamic long-time tails ---} 
In order to understand the origin of the long times required for the onset of the asymptotic diffusive behavior, we compute the velocity autocorrelation function (VACF)~\cite{alder1970decay, van1982transport} 
\begin{equation}\label{3:vacf}
    C(t,t_0)=\langle\mathbf{v}(t_0) \cdot \mathbf{v}(t)\rangle
    \, , 
\end{equation}
where $\mathbf{v}(t)$ is the velocity (namely, the instantaneous direction of motion, given our choice of units for space and time) of the monopole random walk at time $t$. In steady state, it depends only on the time difference: $C=C(t-t_0)$. 
The VACF is directly related to the MSD~\cite{van1982transport}: 
\begin{equation}\label{3:vm}
    C(t)=\frac{1}{2}\frac{\mathrm{d}^2}{\mathrm{d}t^2}M(t)\,;\quad M(t)=\int_0^t \mathrm{d}s\,C(s)(t-s)
    \, .
\end{equation}
In our model, the VACF measures the autocorrelation of the directions of the flipped spins 
(see App.~\ref{app:methods}). 
 
We plot the VACF of the monopole and of the {\SPC} random walks in the bottom right panel of Fig.~\ref{fig:1}. A negative $C(t)$ is consistent with the behavior of the MSD. At long times, we observe a clear scaling $C(t)\sim t^{-3/2}$ over more than a decade in time. From Eq.~\eqref{3:vm}, this suggests a long-time expansion of $M(t)$ of the form
\begin{equation}\label{3:expand1}
    M(t)=Dt+A_1t^{1/2}+{O}(t^0)
    \, , 
\end{equation}
which is confirmed by fitting the MSD at long times ($t>10^2$) in Fig.~\ref{fig:1} (top right panel), with parameters $D\simeq1.85\pm0.01$ and $A_1\simeq0.9\pm0.1$. 

Despite bearing similarities with the monopole random walk at short times, the {\SPC} walk gives $C(t)\sim t^{-5/2}$ at long times, consistent with earlier results on random walks with quenched disorder~\cite{van1982transport, nieuwenhuizen1986diffusion, machta1981generalized} (see also App.~\ref{app:ltt}). 
This explains its faster convergence. In fact, Eq.~\eqref{3:vm} for the {\SPC} walk gives Eq.~\eqref{3:expand1} without the contribution $\propto t^{1/2}$. 
%
%

\emph{Self-reinforcement effects ---} 
The results presented above, in particular the scaling of the VACF and the reduced long-time asymptotic value of the diffusion constant, are indicative of important memory effects at play in the monopole random walk, which are clearly not captured by a {\SPC} model of frozen blocked directions. This calls for a connection to the broader class of random walks characterized by self-reinforcement effects, where the walker at a given time $t$ interacts with its own trail at $t'<t$. 

\begin{figure}[t!]
    \includegraphics[width=0.48\linewidth]{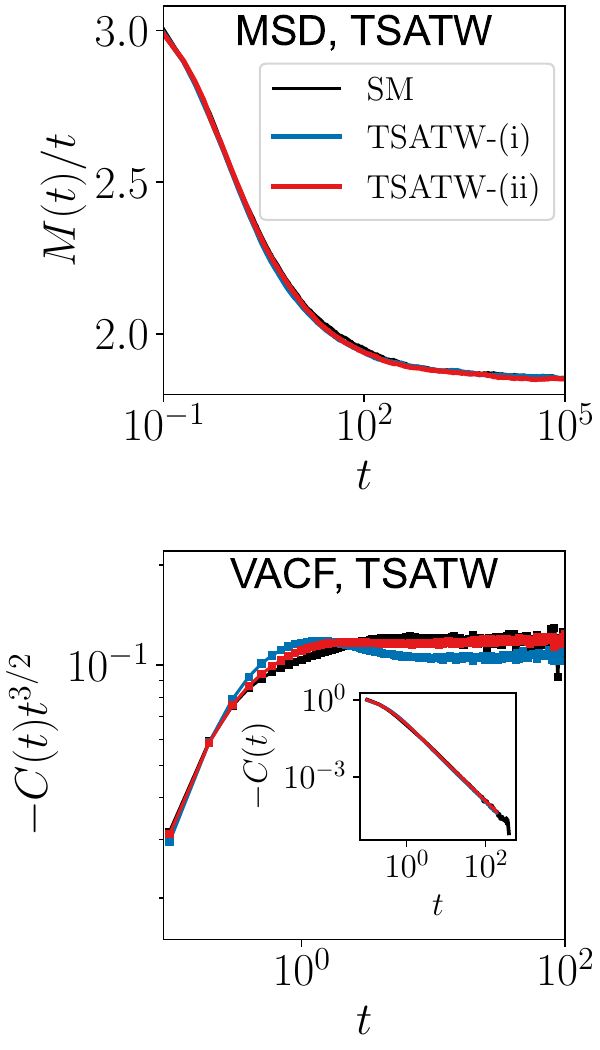}
    \includegraphics[width=0.48\linewidth]{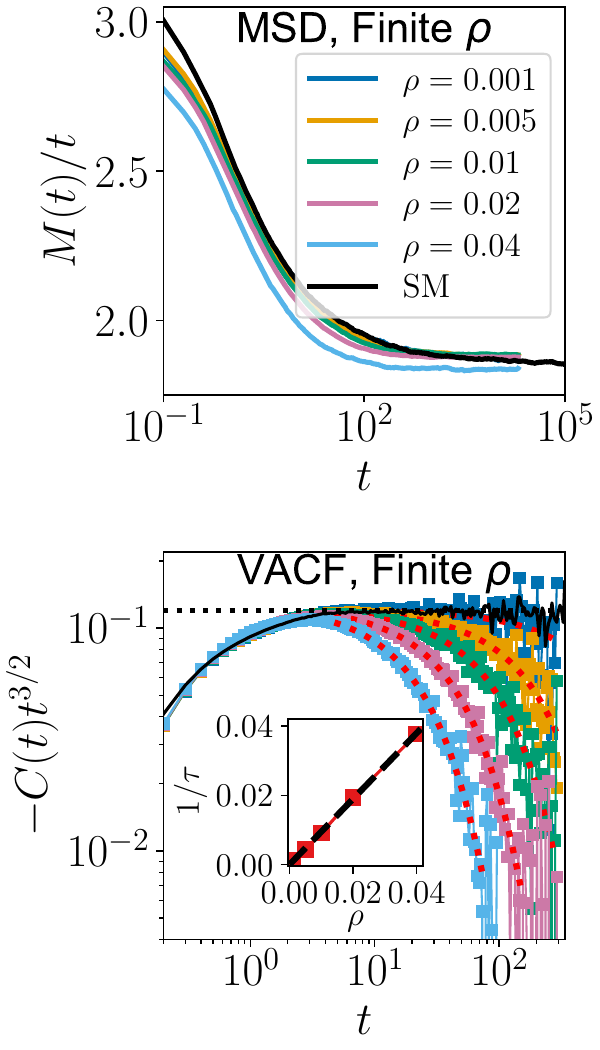}
    \caption{\label{fig:2} Left panels: comparison between the monopole random walk and the TSATW with suitably chosen parameters (see main text). We find good agreement for both the MSD (top) and VACF ($-C(t)t^{3/2}$, bottom main; $-C(t)$, bottom inset). 
    Right panels: simulations of standard model spin ice with fixed finite density of monopoles (microcanonical ensemble). The MSD (top) converges to the diffusive limit more quickly the higher the monopole density, and $-C(t)t^{3/2}$ (bottom) exhibits a long-time exponential decay due to monopole motion altering other monopoles' memory trail (Dirac strings). 
    Fits to the VACFs of the form $C(t)\sim t^{-3/2} e^{-t/\tau}$ are shown as red dotted lines. Bottom right panel inset: monopole density $\rho$ (red squares) versus inverse characteristic decay time, $1/\tau$, and its linear fit $\rho\tau \simeq 0.95\pm0.01$ (black dashed line). 
    }
\end{figure}

For comparison, we consider the so-called `true self-attractive random walk' (TSATW)~\cite{foster2009reinforced, barbier2022self}, where the walker tends to be attracted by its past trail. Motivated by the standard model monopole dynamics, we define, for each oriented nearest-neighbor pair of sites $i$ and $j$, $n_{ij}$ as the number of times that the random walk particle traverses the oriented bond backwards, i.e., from site $j$ to $i$ (counting from some chosen initial time $t=0$). The transition rate from site $i$ to site $j$ is then assumed to be proportional to a monotonically non-decreasing function $w(n_{ij})$, with $w(0)=1$. Based on insights from previous literature (see App.~\ref{app:tsatw}) and using as few adjustable parameters as possible, we propose two forms of $w(n)$ that reproduce reasonably well the monopole random walk MSD and VACF: (i) a piece-wise constant function $w(0)=1$ and $w(n > 0)=5/3$; and (ii) a linear function $w(n)=1+0.5n$. Our interest here is to draw a qualitative comparison, and therefore we choose simple values of the parameters that give good agreement instead of fitting for the best possible values.
In Fig.~\ref{fig:2} (left panels), we plot the MSD and VACF for the TSATW with these two choices for $w(n)$. 
%
%
%

\emph{Finite monopole density ---} 
The single-monopole case considered so far is appropriate for low temperatures and timescales such that the monopole trails span distances shorter than the typical monopole separation. In order to investigate what happens when different monopole trails overlap (relevant for finite temperatures and long times), we consider standard model spin ice systems with finite (fixed, i.e., microcanonical) monopole density. 

On general grounds, we expect the motion of a monopole (and its corresponding spin flips) to scramble the memory of the Dirac strings of other monopoles, thereby disrupting the dynamical processes we identified as responsible for the long-time hydrodynamic behavior. 
The survival rate of the memory left behind by a certain monopole during time $t$ is thus proportional to a Poissonian factor $e^{-t/\tau}$, where $\tau$ is a characteristic timescale that is intuitively inversely proportional to the monopole density. 
Indeed, we observe that the VACF of a monopole acquires an exponential decay on top of the $t^{-3/2}$ scaling, which is well-captured by the modified form $C(t)\sim t^{-3/2} e^{-t/\tau}$ at sufficiently large times (see bottom right panel in Fig.~\ref{fig:2}). In the inset, we plot $1/\tau$ against the monopole density $\rho$, to confirm their proportionality: $\rho \tau = 0.95\pm0.01$ from a linear fit. 
The MSD is correspondingly seen to approach the asymptotic diffusive regime more quickly as the monopole density increases (see top right panel in Fig.~\ref{fig:2}). 

We also simulate the scenario in which all but the observed monopole are kept static (see App.~\ref{app:finite}). In that case, we find the exponential cutoff to be absent and $C(t)$ scales robustly as $t^{-3/2}$ (for the same monopole densities). This clearly indicates that the exponential decay is not due to the sole presence of other monopoles, but to their motion. 
This explicitly rules out other possible origins of the exponential decay, such as collisions between hard-core monopoles or algebraic (dipolar) spin correlations (which are curtailed by a finite density of static monopoles~\cite{udagawa2021spin}). 

Simulations of the TSATW with finite-time memory consistently lead to similar results (cutoff in the VACF and earlier onset of the asymptotic behavior in the MSD), as shown in App.~\ref{app:tsatw}. 
%
%

\begin{figure}
    \centering
    \includegraphics[width=0.7\linewidth]{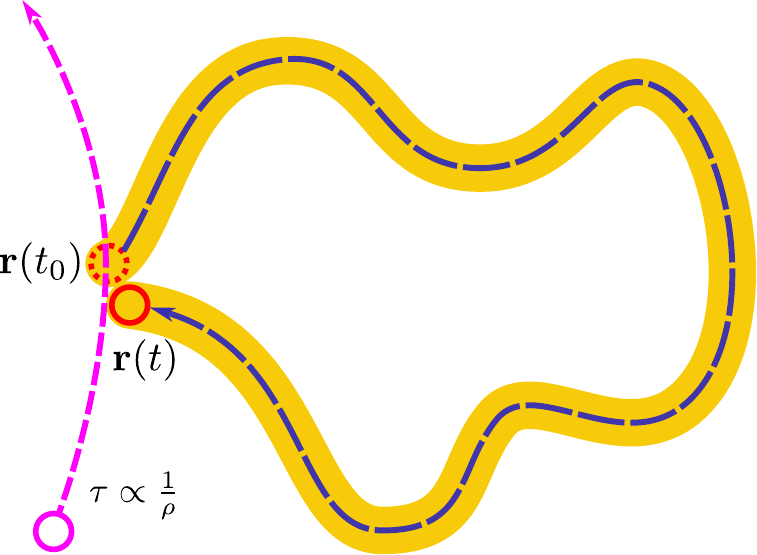}
    \caption{Schematic illustration of the origin of the hydrodynamic memory of a monopole random walk in spin ice. The monopole (red circle) wanders across the system (black dashed line), leaving behind a memory trail of traversed spins (yellow brush) with which it later interacts. At finite monopole density, other monopoles may scramble the stored memory (pink circle and dashed line), with a characteristic timescale inversely proportional to the monopole density.
    }
    \label{fig:monopole}
\end{figure}

\emph{Conclusion --- }
A schematic illustration of the hydrodynamic memory in the standard model monopole random walk is presented in Fig.~\ref{fig:monopole}. Memory is created as a given monopole wanders across the system, leaving behind Dirac strings (i.e., trails of traversed spins) with which the monopole later interacts by intersecting them. 
The VACF $C(t)$ can then be approximated as proportional to the return probability of the walker, which in $d$ dimensions is known to scale as $t^{-d/2}$, provided that the walker is not too strongly constrained (e.g., the MSD should not be bounded). This gives indeed $C(t)\sim t^{-3/2}$ in three dimensions, and is consistent with earlier results on the monopole random walk in $d=2$ square ice~\cite{sutcliffe2022thermal} that found a VACF scaling close to $t^{-1}$ (see App.~\ref{app:2d}). A finite density of monopoles introduces an exponential cutoff to the VACF via erasure of the stored memory, which has also been observed in $d=2$ square ice~\cite{chakraborty2025fractional}. 

Monopole dynamics can be accessed in spin ice experiments via time-dependent magnetization measurements. Particularly fruitful of late have been SQUID measurements of magnetic noise~\cite{watson2019real,dusad2019magnetic, wang2021monopolar,samarakoon2022anomalous,hallen2022dynamical,hsu2024dichotomous}. While these relate directly to the monopole MSD, the hydrodynamic memory effects discussed in our paper are in fact obscured by details of the spin dynamics beyond the standard model considered in this work (see the emergent dynamical fractal in Ref.~\onlinecite{hallen2022dynamical}: random walks on a percolation fractal give $M(t)\sim t^{0.5}$ and thus $C(t)\sim t^{-1.5}$~\cite{eduardo1990diffusion}, indistinguishable from our hydrodynamic memory exponent). 
For completeness, we present a study of the PSD for standard model~\cite{jaubert2009signature} and beyond the standard model~\cite{hallen2022dynamical} dynamics in App.~\ref{app:psd} to illustrate this point, including the effects of dipolar interactions~\cite{udagawa2021spin}. 

Direct probing of the VACF requires a high degree of spatiotemporal resolution, which is highly non-trivial to achieve but may be accessible via state of the art nano-SQUIDs or nitrogen vacancy centers in diamond. Alternatively, present-day quantum simulation platforms allow for analogous spatiotemporally-resolved measurements. 

In conclusion, we have achieved progress in two directions. First, we have identified signatures of memory in the random walks executed by monopoles in spin ice -- succinctly, monopoles move like ants~\cite{dussutour2004optimal}. Our results thus inform future dynamical modeling of spin ice and its variants~\cite{hallen2024thermodynamics, sala2014vacancy, pei2025random}. Second, we have provided a general framework applicable to a broad class of kinetically constrained systems, including dimer~\cite{Moessner2011dimer} and vertex models~\cite{ritort2003glassy}, which can act as a case in point for a wide range of stochastic processes beyond unrestricted Brownian behavior. 
%
%

\section*{Data availability Statement}
All data produced and presented in this work were obtained by well-known methods that have been described in detail in the main text and App.~\ref{app:methods}, and can be readily reproduced. They can also be made available upon reasonable request.
%
%

\section*{Acknowledgements} 
We thank Nilotpal Chakraborty and Santiago A. Grigera for useful discussions, and Attila Szab\'{o} for kindly providing code for dipolar spin ice simulations. 
This work is supported in part by the Engineering and Physical Sciences Research Council (EPSRC) grant No.~EP/V062654/1 and the Deutsche Forschungsgemeinschaft via the cluster of excellence ctd.qmat (EXC 2147, project-id 390858490) and  FOR 5522 (Project-ID No. 499180199).
%
%

\appendix

\section{Methods\label{app:methods}}
In our single-monopole simulations, we initialize a random spin ice system of size $l$ -- that is, $l\times l\times l$ unit cells, each containing $16$ spins (see left bottom panel of Fig.~\ref{fig:1} in the main text) -- with periodic boundary conditions, and let it evolve at $T=0$ until no monopoles are present. The system is then enlarged by stacking $m$ copies in each direction, yielding a configuration of size $L=ml$ that fulfills the ice rule everywhere. Loop updates are applied to decorrelate the enlarged system. In each update, a monopole-antimonopole pair is created by flipping a random spin; we fix one and let the other wander in the system by successive spin-flip attempts of the four adjacent spins, until it meets again the other monopole and they annihilate, creating a loop of updated spins. To facilitate the process, we disfavor short loops and trivial backtracking by forbidding annihilation for a time interval of $2\times10^5$. 
We carry out the simulation and compute the desired quantities using $\delta t = 0.1$ throughout this work, meaning that an attempted flip of one of the four spins adjacent to a monopole occurs with probability $0.1$ at each timestep, while the algorithm has a probability $0.6$ of not attempting any spin flip within a single timestep. 

To obtain the MSD of the spin ice monopole random walk, we set $l=30$, $m=6$, therefore $L=180$. With such a system size, $10$ loop updates can reduce the autocorrelation function between the initial and final states to approximately $10^{-2}$, and we chose to perform $100$ loop updates to prepare the system.
In each simulation, we first create a monopole-antimonopole pair and fix one of them. We then let the mobile monopole wander for a time $t_{\mathrm{eq}}=2\times10^5$ before measuring its MSD by monitoring its position for another $t=t_{\mathrm{eq}}$, preventing annihilation events throughout the entire process. This process is repeated $20$ times before we let the mobile monopole continue to wander until it rejoins its antimonopole, and a new ice-rule-satisfying configuration is generated by annihilating them. This continuously generates decorrelated initial states throughout the simulation. Each dataset presented in the paper is averaged over $4\times10^6$ histories (distinct Monte Carlo simulations).

To obtain the VACF, we use the same system size and loop update scheme. As mentioned in the main text, the VACF is equal to the correlation between spin flip directions. In particular, let $\hat{\mathbf{e}}(t)=\mathbf{0}$ if no spin is flipped at time $t=0, \delta t, 2\delta t, \dots$, and let it be the unit vector parallel to the flipped spin otherwise; then, we define 
\begin{equation}\label{1:vacf}
    C(t)=\langle\hat{\mathbf{e}}(0)\cdot\hat{\mathbf{e}}(t)\rangle
    = \frac{1}{(t_{\mathrm{max}}-t)\delta t^2}\sum_{\tau=0}^{t_{\mathrm{max}}-t}\hat{\mathbf{e}}(\tau)\cdot\hat{\mathbf{e}}(t+\tau)\,,
\end{equation} 
where the factor $\delta t^2$ in the denominator ensures that the resulting $C(t)$ is invariant under the particular choice of $\delta t$. 
The VACF curves in the paper are averaged over $8\times10^5$ trajectories. 

In our simulations of the {\SPC} random walk, we first generate a diamond lattice of size $L=20$ with an ordered dimer pattern, which is then decorrelated by applying $5\times10^4$ loop updates~\cite{Moessner2011dimer}. We perform $20$ random walk simulations on such a configuration, after which the dimer pattern is renewed by applying again $5\times10^4$ loop updates. We repeat this for a total of $2\times10^5$ histories to produce the data presented in this paper. We verified that doubling the system size to $L=40$ does not induce any noticeable changes in the measured quantities.
%
%

In our simulation of the TSATW, similar to the monopole random walk, we take $\delta t=0.1$. In each simulation, we equilibrate the system over $2\times10^5$ time units before making any measurements to minimize transient effects. Each of the datasets presented in this paper is obtained by averaging over $4\times10^5$ Monte Carlo trajectories (for details on the TSATW rules, see App.~\ref{app:tsatw}.). 
%
%

To obtain the MSD and VACF results in the presence of a finite density of monopoles (right panels of Fig.~\ref{fig:2} in the main text), we simulated spin ice systems of size $L\times L\times L$ using the conventional single-spin-flip Monte Carlo method with Metropolis-Hastings acceptance probability: spins are randomly drawn and flipped with a probability $p=\min\left(1,\,\mathrm{e}^{-\beta \Delta E}\right)$, with $\Delta E$ being the energy difference caused by the flip. We identify $N_s=16L^3$ such attempts as one Monte Carlo time step. 
To achieve the desired monopole density, we start with a random spin configuration and gradually lower the temperature until the desired monopole density is achieved. We then forbid monopole creation or annihilation events and run our simulations in the microcanonical ensemble for $N_{\mathrm{sim}}=2\times10^5$ steps, during which we keep track of the position of every monopole. The MSDs and VACFs are then computed as usual. We studied systems of size $L=15$ to $40$, with $L$ smaller for larger $\rho$. 
%
%

\section{Long-time tails\label{app:ltt}} 
The concept of long-time tails refers to the asymptotic algebraic decay of dynamical correlation functions. 
In the standard Langevin description of Brownian motion, the VACF decays exponentially, reflecting a short memory kernel~\cite{kubo2012statistical}. Alder and Wainwright instead found that in molecular-dynamics simulations of hard-sphere fluids the VACF decays algebraically, with the asymptotic form $C(t)\sim t^{-d/2}$ in $d$ dimensions~\cite{alder1970decay}. This slow relaxation originates from the feedback of long-lived hydrodynamic modes of the surrounding fluid onto the tagged particle.  Subsequent work established long-time tails as a general consequence of mode coupling and hydrodynamic memory, and showed how they may be described by generalized Langevin equations with memory kernels replacing strictly Markovian friction~\cite{ernst1971asymptotic,pomeau1975time,paul1981observation,van1982transport,fox1983long,ferrario1997long}. Long-time tails have since been discussed in a wide range of systems, including quantum many-body dynamics~\cite{lux2014hydrodynamic}, cellular automata~\cite{nava2017cellular,naitoh1990long}, and field-theoretic settings~\cite{kovtun2003hydrodynamic}. 

It is useful to contrast this hydrodynamic-memory mechanism with long-time tails generated by static spatial disorder. A paradigmatic example is the Lorentz gas, where a light particle scatters from fixed obstacles. In this case the VACF exhibits a different asymptotic form, $C(t)\sim t^{-(1+d/2)}$ for $d\leq 3$, reflecting repeated encounters with a quenched environment rather than feedback from a dynamically evolving medium~\cite{van1982transport,machta1981generalized,zwanzig1982non}. Related behavior occurs in Lorentz lattice models and in random walks on percolation clusters~\cite{nieuwenhuizen1986diffusion}. The structured-percolation-cluster walk considered in our work belongs to this broad class of walks in static constrained environments: in both $d=3$ and $d=2$ it exhibits the corresponding $C(t)\sim t^{-(1+d/2)}$ behavior, giving $C(t)\sim t^{-5/2}$ in three dimensions and $C(t)\sim t^{-2}$ in two dimensions. 
%
%

\section{Two-dimensional square spin ice\label{app:2d}}
\begin{figure*}[ht]
    \centering
    \includegraphics[width=0.36\linewidth]{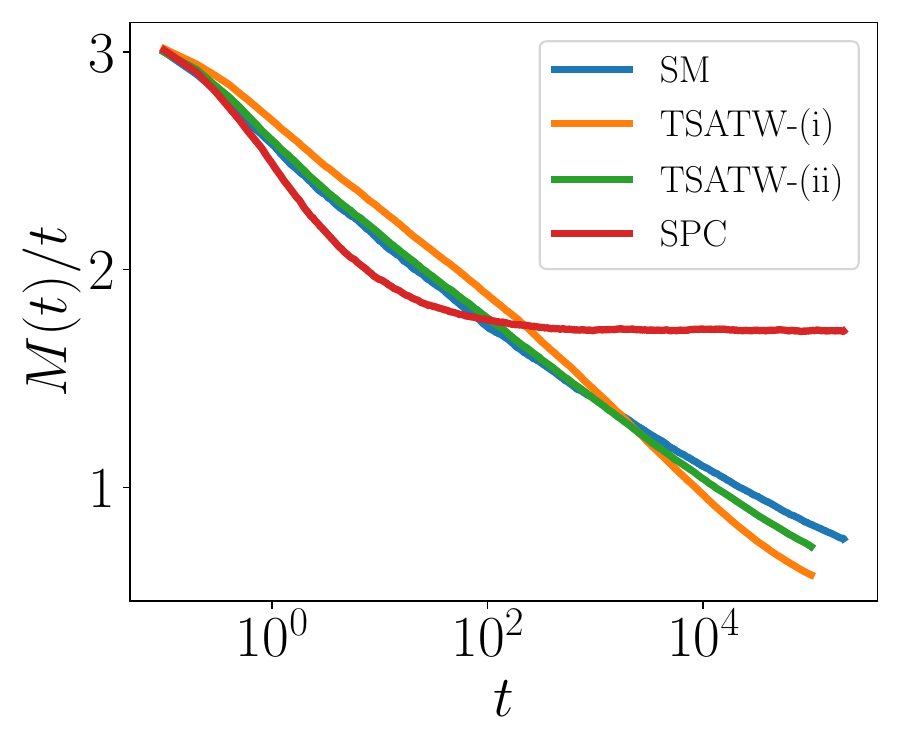}
    \includegraphics[width=0.36\linewidth]{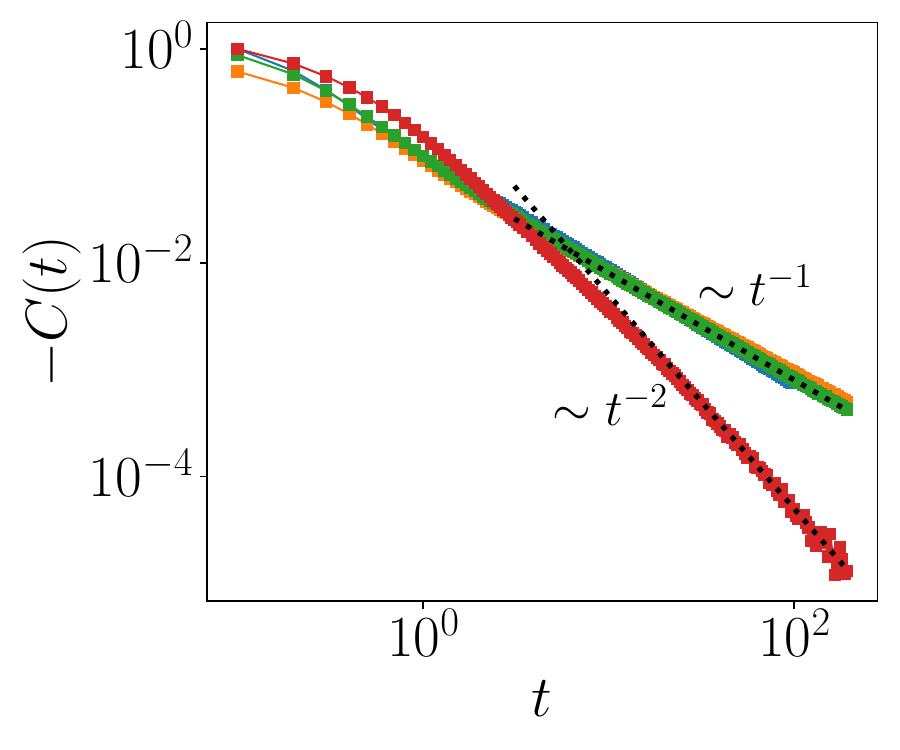}
    \caption{Two-dimensional square spin ice results. (a) MSD divided by time, $M(t)/t$, for the monopole random walk in $d=2$ and for two TSATWs on the square lattice (see text); we also present results for the {\SPC} random walk on the square lattice. The VACFs $-C(t)$ for the same processes are shown in panel (b).} 
    \label{fig:1:2d}
\end{figure*}
We present here results for two-dimensional square ice for comparison, referring the reader to existing literature for further details~\cite{sutcliffe2022thermal, chakraborty2025fractional}. 

The MSD and VACF of the monopole random walk in $d=2$ are shown in Fig.~\ref{fig:1:2d}. We carried out simulations 
similarly to our $d=3$ study, by merging $l=200$ ice-rule-satisfying configurations in a $6\times6$ grid (hence forming a system of linear size $L=1200$) and carrying out $100$ loop updates to decorrelate it. In each simulation, we ran for time $2\times10^{5}$ for both equilibration and measurement, and we averaged our data over $4\times10^6$ histories. 

We also present results for the TSATW and the {\SPC} random walk on the square lattice. The TSATW rules are identical to those formulated in the main text, with $w(n)$ modified to be either (i) $w(0)=1$ and $w(n \ge 1)=2.5$, or (ii) $w(n)=1+0.5n$. The {\SPC} random walk is simulated on a percolation cluster where the removed bonds form a dimer configuration on an $L=200$ square lattice. 

The large-$t$ behavior of the monopole random walk VACF is consistent with $t^{-1}$ decay, expected from our prediction of $t^{-d/2}$ for $d=2$ (see Ref.~\onlinecite{sutcliffe2022thermal} for a detailed analysis of finite size corrections). 
This behavior is also observed in the TSATWs that we chose to study in our work. 

%
%

\section{Beyond the standard model\label{app:bSM}}
\begin{figure*}
    \centering

    \subfloat	[\label{fig:bSM(a)}]{
        \includegraphics[width=0.27\linewidth]{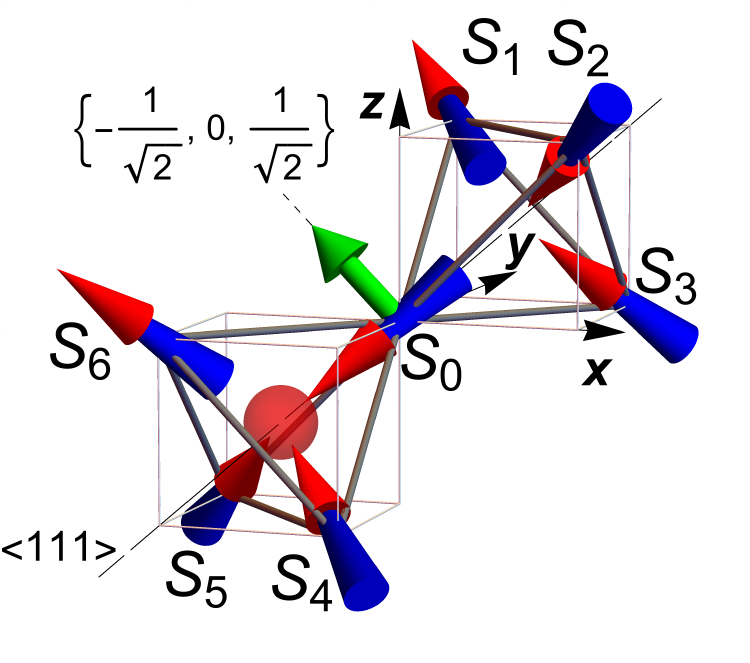}
    }
    \subfloat	[\label{fig:bSM(b)}]{
        \includegraphics[width=0.27\linewidth]{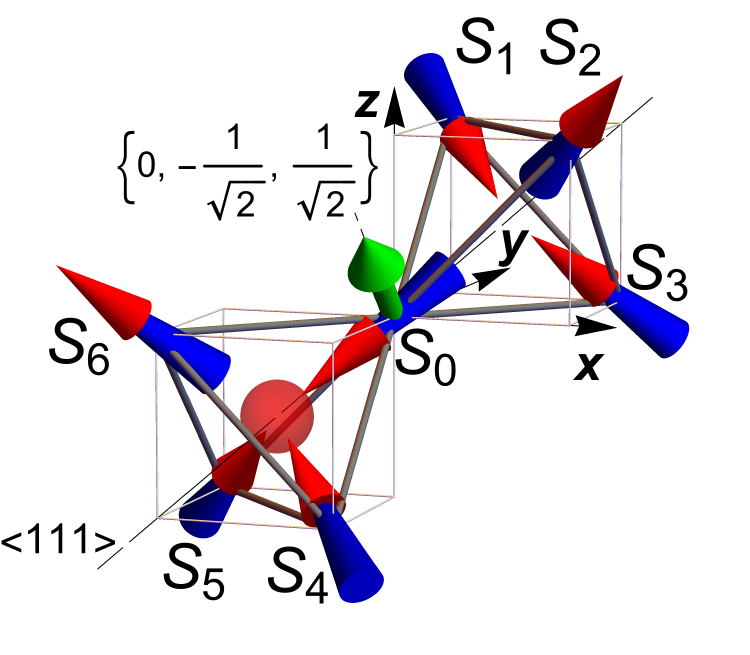}
    }
    \subfloat	[\label{fig:bSM(c)}]{
        \includegraphics[width=0.27\linewidth]{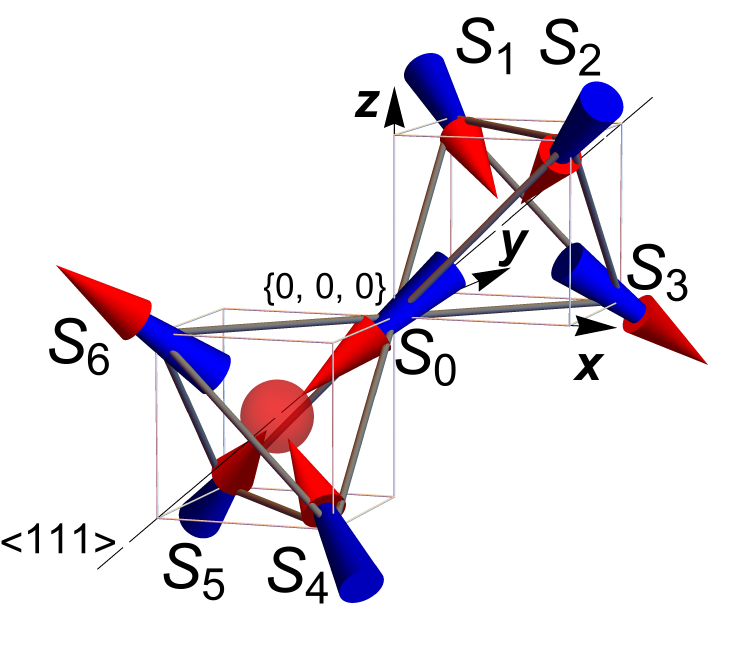}
    }
    
    \subfloat	[\label{fig:bSM(d)}]{
        \includegraphics[width=0.27\linewidth]{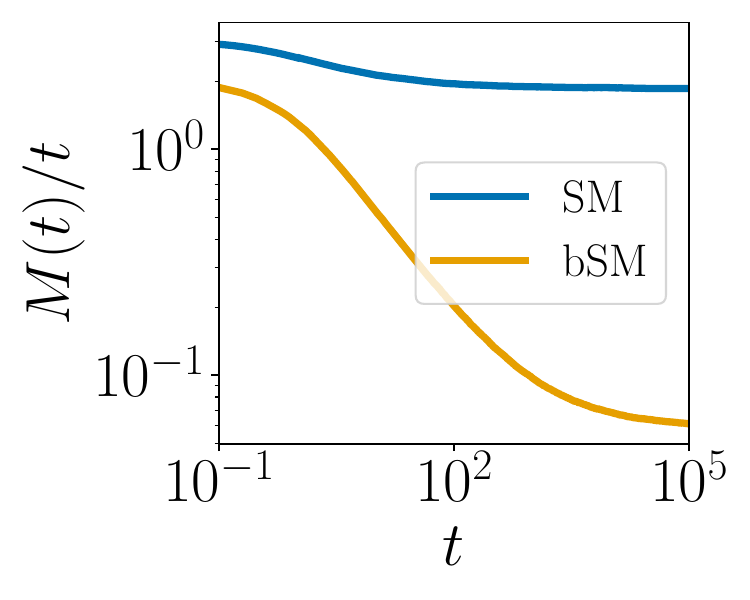}
        }
    \subfloat	[\label{fig:bSM(e)}]{
        \includegraphics[width=0.27\linewidth]{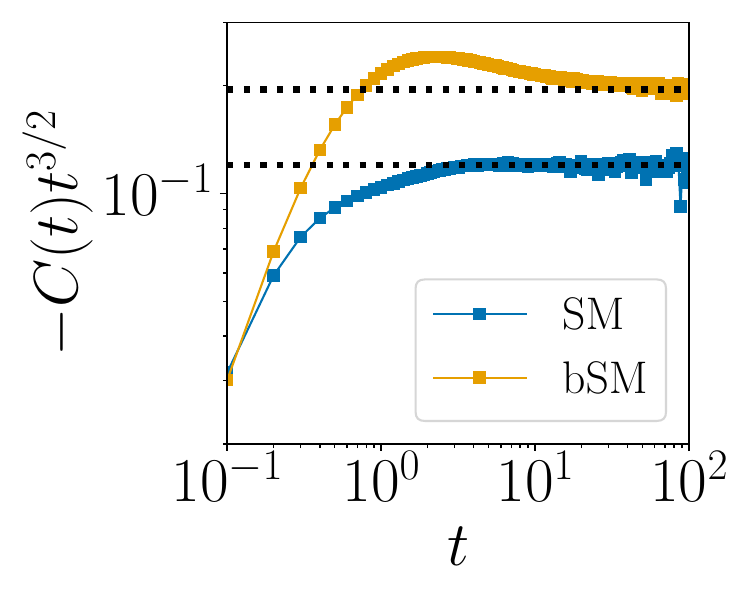}
        }
    \caption{Illustration of symmetry-inequivalent configurations for a single spin and its nearest neighbors (a)-(c), for the case where flipping the central spin leads to monopole hopping. The direction of the transverse field due to nearest-neighbor exchange interactions is shown as a green arrow; it vanishes exactly by symmetry, and therefore it is not shown, in panel (c). We present a comparison between standard model (SM) and bSM results in panels (d), MSD divided by time $M(t)/t$, and (e), $-C(t)t^{3/2}$. The horizontal dashed lines are guides to the eye. Panels (a)-(c) are reproduced from Ref.~\onlinecite{tomasello2019correlated}.}
    \label{fig:S:bSM}
\end{figure*}

We present here a brief review of the spin ice dynamics dubbed `beyond the Standard Model' (bSM)~\cite{hallen2022dynamical}. It stems from a microscopic understanding of spin flips in spin ice materials as quantum tunneling events driven by the transverse component of the local exchange interactions.
Considering for simplicity only nearest neighbor contributions, spin flips leading to monopole motion can have $3$ symmetry-distinct neighboring spins configurations, shown in Fig.~\ref{fig:bSM(a)} to~\ref{fig:bSM(c)}. Among them, one (Fig.~\ref{fig:bSM(c)}) has precisely vanishing transverse interaction on the central spin, 
while the others are finite~\cite{tomasello2019correlated, NoteS1}.
The bSM correspondingly associates all spin flips with a vanishing transverse interaction to a vanishing classical stochastic Monte Carlo flipping timescale, 
thereby leading to additional blocked directions~\cite{hallen2022dynamical}. 
At each step of the bSM random walk, there are between $1$ and $3$ active directions (i.e., flippable spins), depending on the local spin correlations (and backtracking remains always possible). 

We plot results for the MSD in Fig.~\ref{fig:bSM(d)} and for the VACF in Fig.~\ref{fig:bSM(e)}, contrasting the single monopole SM and bSM random walks. Naively, one expects on average $2$ active directions per bSM step, as one third of the spins active in SM are blocked in bSM, on average; from our simulation, however, we obtain $\lim_{t\to0}M(t)/t\simeq1.9$, possibly due to the correlated nature of the blocked spins. The bSM demonstrates an anomalous MSD~\cite{bouchaud1990anomalous}, not converging to the diffusive limit even for times as large as $t\sim10^5$, with an estimated asymptotic diffusion constant $D<0.05$. The VACF appears to asymptote to $C(t)\sim t^{-3/2}$ at long times, with a value approximately twice as large as the SM. 
%
%

\section{Power spectral density\label{app:psd}}
For completeness, we present here a study of the power spectral density (PSD) of the magnetic noise of the different models considered in our work. 
From the time-dependent global magnetization of the system, $\mathbf{m}(t)$, one can compute 
\begin{equation}
    \mathrm{PSD}(\omega) \equiv \left\langle\left|\int_{-\infty}^{\infty}\mathrm{d}t\,\mathbf{m}(t)\mathrm{e}^{-\mathrm{i}\omega t}\right|^2\right\rangle
    \, , 
\end{equation}
using, for example, Welch's method~\cite{welch1967use}. 

When the magnetization dynamics derives from the motion of an individual random walker, the PSD is related to the corresponding MSD via  
\begin{equation}\label{5:msdpsd}
    \mathrm{PSD}(\omega)=\frac{1}{\omega}\int_0^\infty \mathrm{d}t\,\sin(\omega t)\,\dot{M}(t)
    \, .
\end{equation}
If the monopole undergoes purely diffusive random-walk dynamics characterized by $M(t)\propto t$, the resulting PSD 
scales as $\mathrm{PSD}(\omega)\propto\omega^{-2}$. Monopole creation and annihilation processes modify the low-frequency behavior, leading to a saturation of the PSD. In this case, the spectrum acquires a Cole-Cole form,  
\[
\mathrm{PSD}(\omega)\propto\frac{1}{1+(\omega\tau_0)^{2}} \, ,
\]  
where 
$\tau_0$ denotes the characteristic timescale associated with spin flip events. 

By contrast, in spin ice with standard model dynamics, Eq.~\eqref{3:expand1} in the main text implies a modified PSD form. In the single-monopole, low-frequency limit (corresponding to the long-time limit of the MSD), one obtains
\[
\mathrm{PSD}(\omega)\propto \omega^{-2}\left[1+(\omega\tau)^{1/2}\right] 
\, , 
\]
where $\tau$ denotes the characteristic timescale over which isolated monopoles recover diffusive behavior. For finite monopole densities, this scaling is expected to manifest at intermediate frequencies, before the onset of the lowest frequency PSD plateau. 

\begin{figure*}
    \centering
    \subfloat	[\label{fig:nnSISM}]{
        \includegraphics[width=0.3\linewidth]{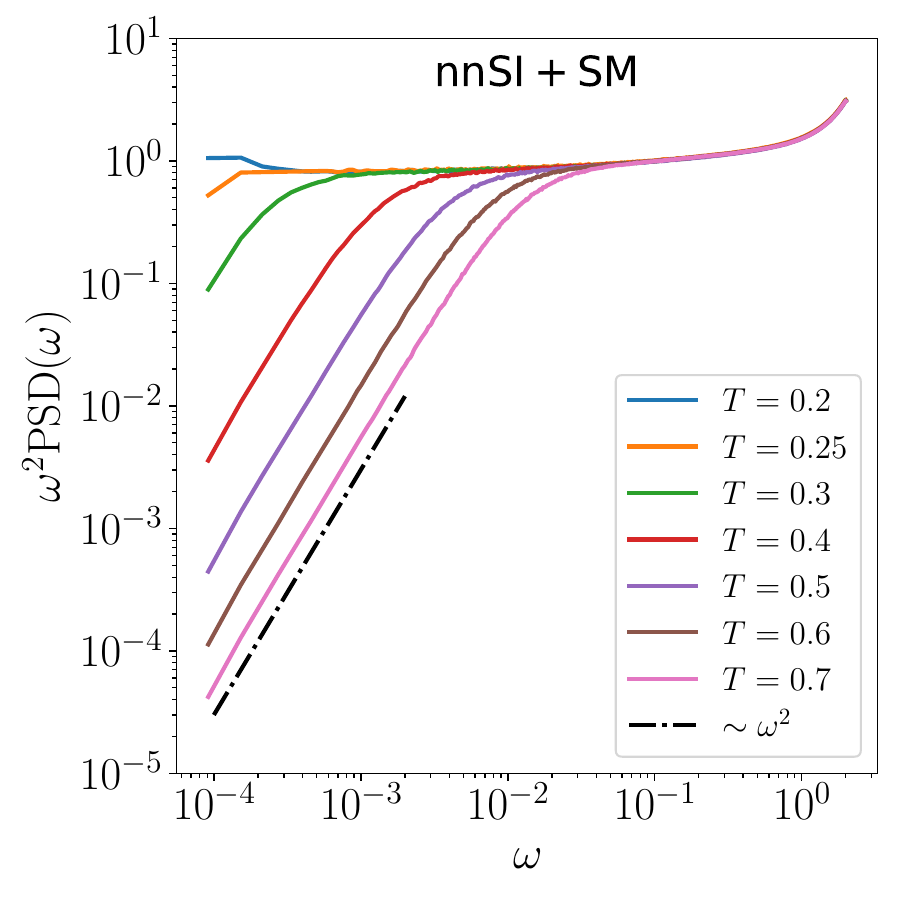}
        }
    \subfloat	[\label{fig:nnSIbSM}]{
        \includegraphics[width=0.3\linewidth]{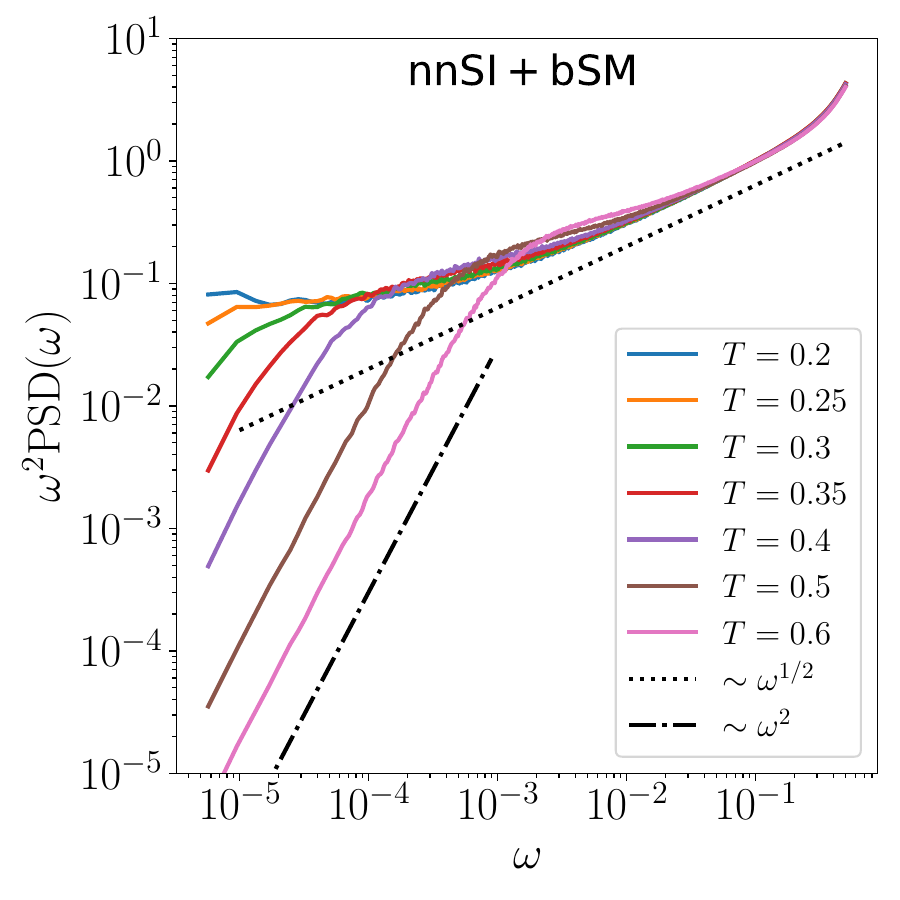}
        }
    \subfloat	[\label{fig:dipSI}]{
        \includegraphics[width=0.3\linewidth]{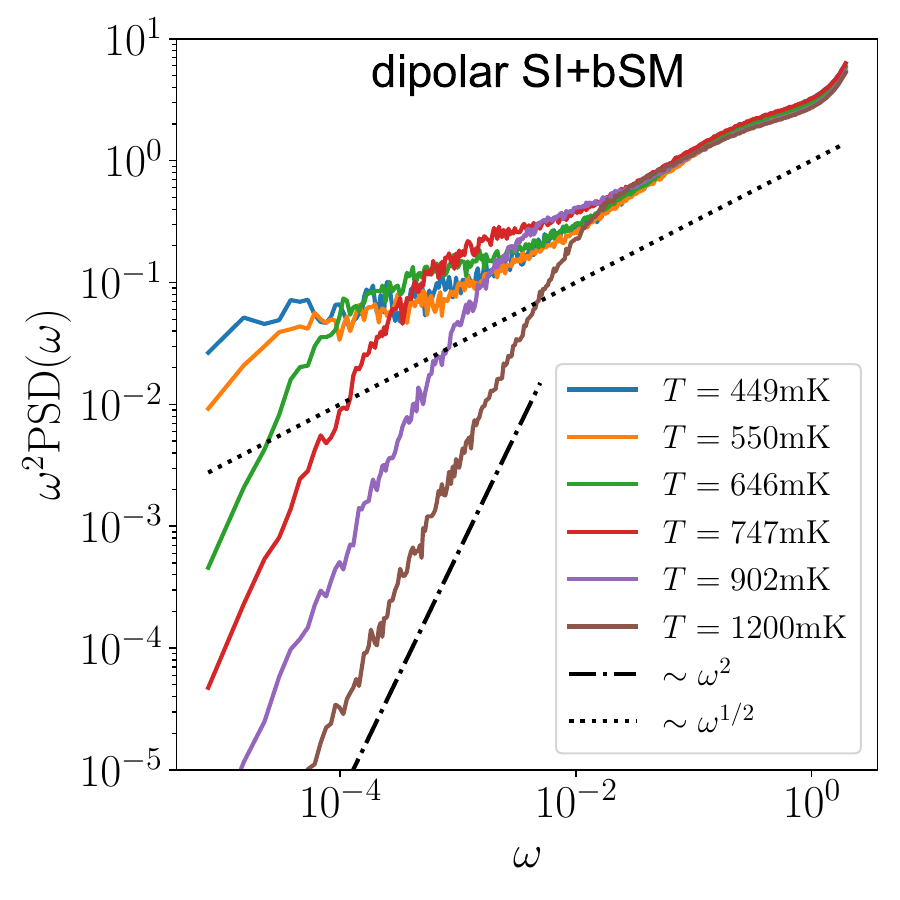}
        }
    \caption{Power spectral density of nearest-neighbor spin ice with SM dynamics (a); nearest-neighbor spin ice with bSM dynamics (b); and dipolar spin ice with bSM dynamics (c). The temperatures are in units of nearest-neighbor coupling in (a) and (b) and in Kelvin in (c). The high-frequency upturn of the PSD is an aliasing effect of Welch's method. 
    In all the panels, the curves have been rescaled by their value at $\omega = 10^{-1}$ for visual clarity, and the frequency axis is in inverse MC time units.}
    \label{fig:S:PSD}
\end{figure*}

In practice, however, the corresponding correction is too weak to be resolved (see Fig.~\ref{fig:nnSISM}): $\mathrm{PSD}(\omega)$ follows a scaling law very close to $\omega^{-2}$ (equivalently, $\omega^{2}\mathrm{PSD}(\omega)$ is approximately constant) until it eventually saturates due to monopole creation-annihilation events (equivalently, $\omega^{2}\mathrm{PSD}(\omega)\sim\omega^{2}$). 

By contrast, in the bSM dynamics a clear crossover is observed from $\omega^{-3/2}$ to $\omega^{-2}$ in the PSD (or, equivalently, from $\omega^{1/2}$ to a constant in $\omega^{2}\mathrm{PSD}(\omega)$, shown in Fig.~\ref{fig:nnSIbSM}) before the spectrum ultimately saturates at low frequencies. This is due to the emergent dynamical fractal from the additional non-flipping spins discussed above~\cite{hallen2022dynamical} rather than from the hydrodynamic memory discussed in this work. 

We also ran simulations of dipolar spin ice, with the Hamiltonian given by
\begin{equation}\label{dipH}
\begin{aligned}
    \mathcal{H}_{\mathrm{dip}} &= D
    \sum_{i>j} 
    \left[
    \frac{\mathbf{S}_i \cdot \mathbf{S}_j}{r_{ij}^3}
    - 
    \frac{3\, (\mathbf{S}_i \cdot \mathbf{r}_{ij})(\mathbf{S}_j \cdot \mathbf{r}_{ij})}{r_{ij}^5}
    \right] \\
    &+ J_1 \sum_{\langle i,j \rangle} \mathbf{S}_i \cdot \mathbf{S}_j 
    + \ldots 
    \, ,
\end{aligned}
\end{equation}
where $\mathbf{S}_i = S_i \hat{\mathbf{e}}_i$; $S_i=\pm1$ are the same Ising spins as in Eq.~\eqref{2:H}; $\hat{\mathbf{e}}_i$ are versors of the local easy axes; and the ellipses denote possible farther range exchange interaction terms (ignored here). We shall use dimensionful parameters for Dy$_2$Ti$_2$O$_7$: $D=1.32$~K and $J_1=-3.41$~K~\cite{samarakoon2020machine}.
We simulated the PSD of dipolar spin ice and bSM dynamics, using the Ewald summation method~\cite{de1980simulation} to account for the long-range dipolar interactions, see Fig.~\ref{fig:dipSI}. We observe a behavior somewhat similar to the nearest-neighbor bSM spin ice case. 
%
%

\section{TSATW\label{app:tsatw}}
\begin{figure*}
    \centering
    \subfloat	[\label{fig:siteTSATW}]{
        \includegraphics[width=0.24\linewidth]{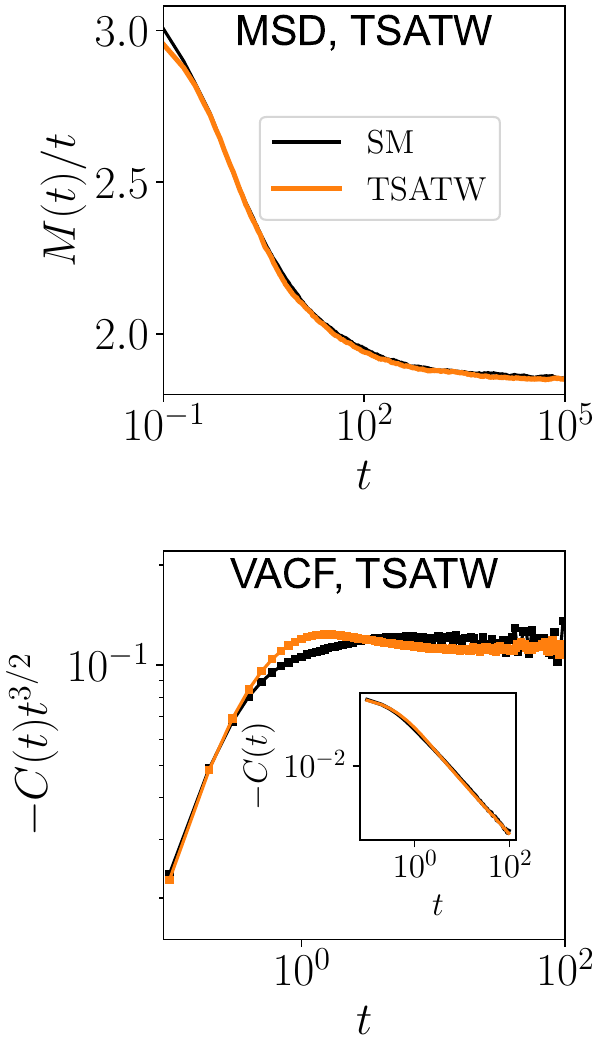}
        }
    \raisebox{3.3em}{
        \subfloat	[\label{fig:fmTSATW}]{
            \includegraphics[width=0.64\linewidth]{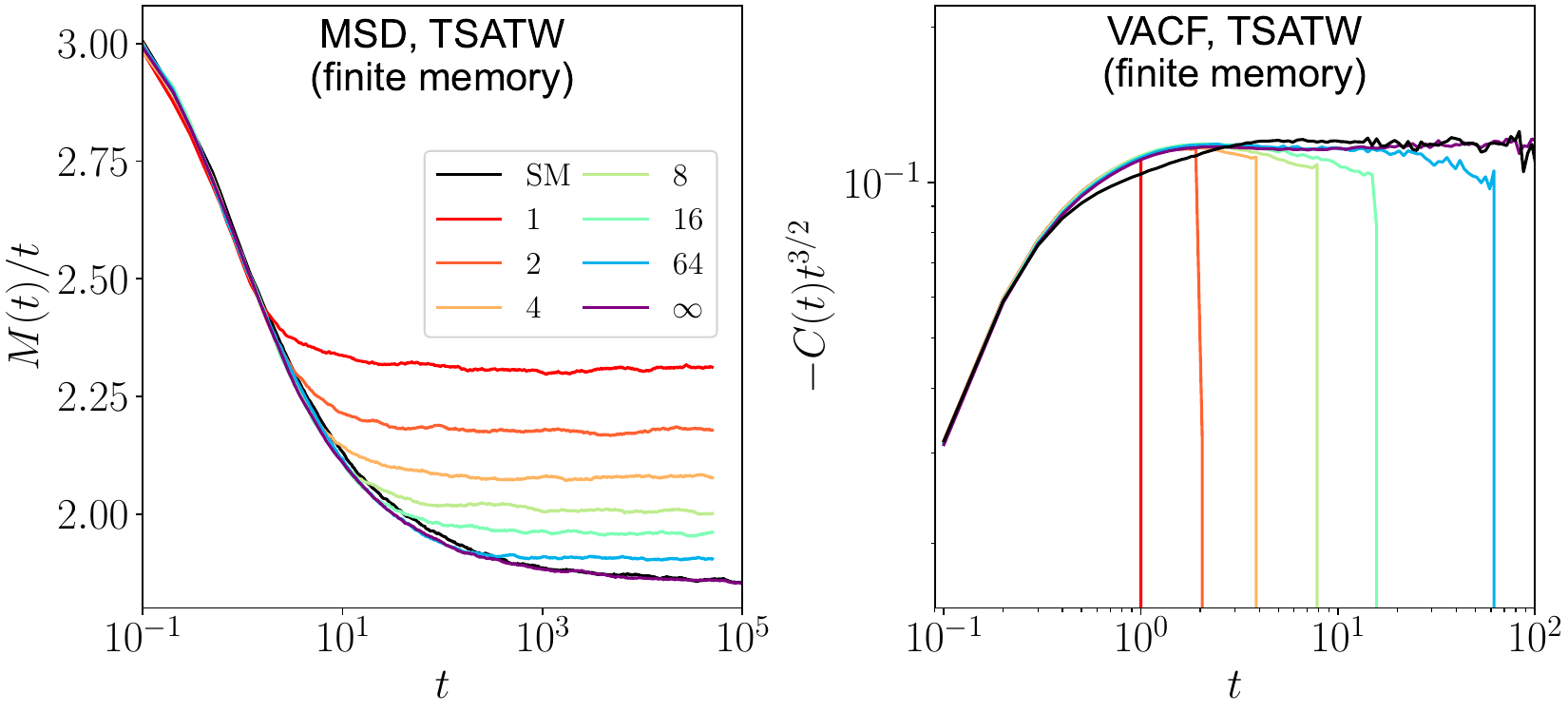}
        }
    }
    \caption{TSATW-(i) with $w(n)=1$ for $n\le 3$ and $w(n)=3.3$ for $n>3$ (a). Finite-memory TSATW with $w(n)=1+0.5n$ (b), where the legend indicates the memory window $\Delta t = 1,2,4,8,16,64,\infty$, the latter value being the case for regular (infinite-memory) TSATW.} 
    \label{fig:S:tsatw}
\end{figure*}
\begin{figure*}[htbp]
    \centering
    \includegraphics[width=0.6\linewidth]{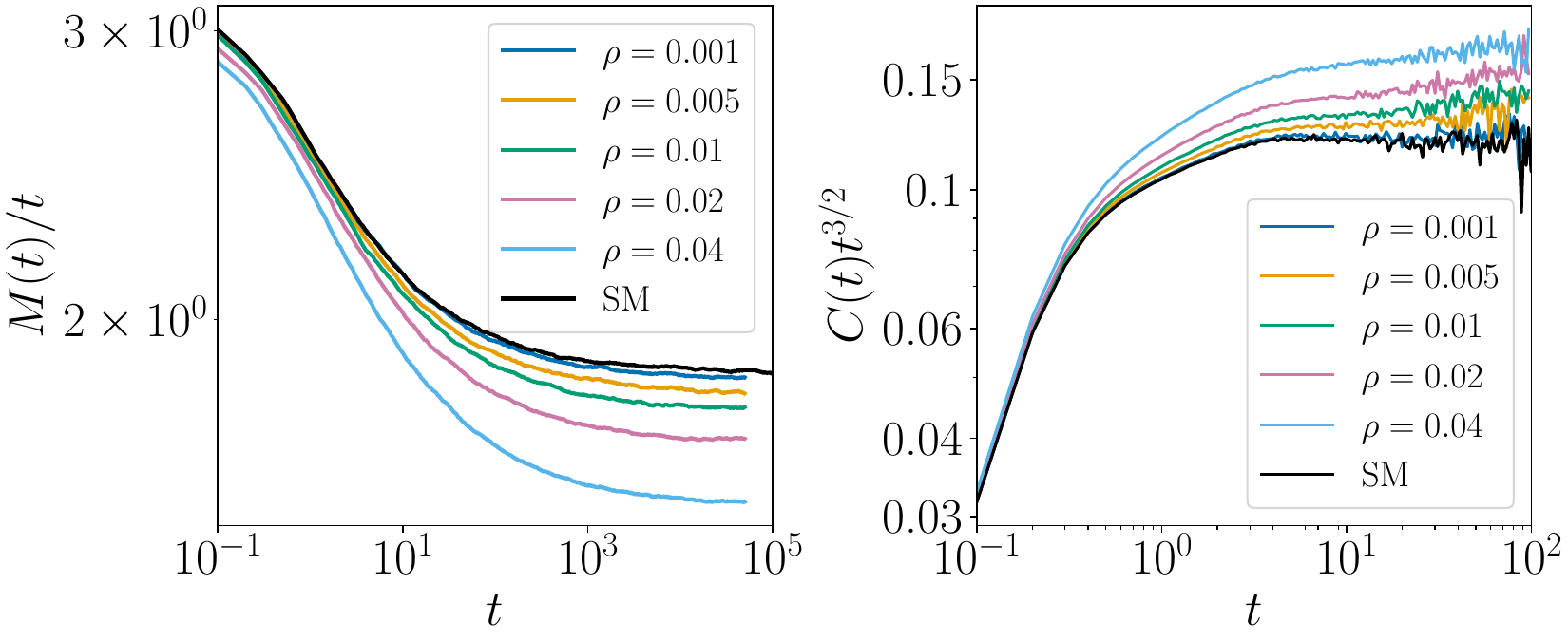}
    \caption{MSD and VACF results for systems with finite monopole density, when all but one monopole are held at fixed positions: $M(t)/t$ (a) and $C(t)t^{3/2}$ (b).} 
    \label{fig:S:FT}
\end{figure*}
The true self-attractive random walk (TSATW) belongs to the broader class of self-interacting random walks, namely random walks on regular lattices where the transition probability from a site $i$ to a nearest-neighbor site $j$ is proportional to some weight function, $w(n)$, where $n$ is a counter of previous visits. In previous literature, $n$ is usually set to count either (i) the number of visits to site $j$ or (ii) the number of times the bond $ij$ has been traversed. In the main text, we use a slightly modified rule (iii), where $n$ counts the number of directed hops from site $j$ to $i$.

Using case (i) as our working example, the transition rate from site $i$ to $j$ is given by
\begin{equation}\label{eq:tsatw}
    P(i\to j)=\frac{w(n_j)}{\sum_{k\in\mathrm{nn}}w(n_k)}\,,
\end{equation}
with $k$ enumerating all nearest-neighbor sites of site $i$. The transition rates for the other cases can be defined similarly.

In three dimensions, it was shown~\cite{barbier2022self} that the random walk is bounded (i.e., bounded MSD) for any $w(n)$ growing exponentially in some power of the system size. When instead $w(n)$ saturates, e.g., $w(n)=w$ for $n>1$, it was shown that there exists a threshold value $w_c\approx6.24$ that separates diffusive and sub-diffusive behavior -- with the same result also applying to case (ii), resulting in a larger $w_c\approx11.88$~\cite{foster2009reinforced}. These earlier results informed our intuition to find a form of $w(n)$ that best reproduces the MSD of the standard model monopole random walk: either a bounded form with the bound much less than $w_c$, or an unbounded form with sufficiently slow growth rate. We opted to present results obtained from a suitably bounded and a linearly growing $w(n)$, with rule (iii) applied. 

For completeness, we also studied the TSATW with rule (i), which is common in the literature. Specifically, we chose $w(n)=1$ for $n\le3$ and $w(n)=3.3$ for $n>3$. The results are shown in Fig.~\ref{fig:siteTSATW} and are once again in good agreement with the standard model monopole random walk behavior (not presented in the main text for conciseness). 

It is interesting to simulate a TSATW with finite memory, using for concreteness $w(n)=1+0.5n$ corresponding to TSATW-(ii) in the main text. Here, $n$ is the number of visits within a finite time window $\Delta t$ preceding the hopping time $t$ (and the usual TSATW corresponds to $\Delta t \to \infty$). The results are plotted in Fig.~\ref{fig:fmTSATW}. A sudden drop in the VACF appears for times larger than $\Delta t$, triggered by the corresponding loss of memory. This in turn gives rise to early convergence of the MSD to the diffusive limit. The results are consistent with our finite-density monopole results presented in the main text, where an exponentially decaying VACF arises from a Poissonian process of memory loss due to other mobile monopoles scrambling the Dirac string. 
%
%

\section{Finite static monopole density\label{app:finite}}
As mentioned in the main text, it is interesting to contrast the behavior of the system at finite monopole density $\rho$ between the conventional microcanonical case, and the artificial case where all but one monopole are held at fixed positions (averaged as one would with quenched disorder). 
The MSD and VACF results are plotted in Fig.~\ref{fig:S:FT}. The VACF demonstrates a robust $C(t)\sim t^{-3/2}$ scaling, devoid of any exponential cutoff. The MSD decreases as $\rho$ increases, but it does not demonstrate early convergence, in contrast to the conventional microcanonical case. Such modifications to the MSD can be therefore attributed to the hardcore excluded-volume effect from other static monopoles, which, as demonstrated above, only induce sub-leading contributions to the VACF at low densities. 

The fixed monopoles results, when compared to the microcanonical results presented in the main text, demonstrate that the VACF scaling in spin ice is not a result of the algebraic (dipolar) spin correlations, as the latter are equally curtailed by a finite density of mobile or static monopoles~\cite{udagawa2021spin}. 
%
%

\bibliographystyle{apsrev4-2}
\bibliography{ref}

\end{document}